\documentclass{article}

\newcommand{\be}{\begin{equation}}
\newcommand{\ee}{\end{equation}}
\newcommand{\beqa}{\begin{eqnarray}}
\newcommand{\eeqa}{\end{eqnarray}}
\newcommand{\bean}{\begin{eqnarray*}}
\newcommand{\eean}{\end{eqnarray*}}

\def\thebibliography#1{\section*{REFERENCES\markboth
        {REFERENCES}{REFERENCES}}\list
        {[\arabic{enumi}]}{\settowidth\labelwidth{[#1]}\leftmargin\labelwidth
        \advance\leftmargin\labelsep
        \usecounter{enumi}}
        \def\newblock{\hskip .11em plus .33em minus -.07em}
        \sloppy
        \sfcode`\.=1000\relax}

\begin{document}

\title{
\begin{flushright}
\baselineskip=11pt
{\small ESI--1110}\\ 
{\small gr--qc/0202018}\\
\end{flushright}
\vspace{1.5cm}
On the $SO(2,1)$ symmetry in General Relativity
}
\author{Gaetano Vilasi and Patrizia Vitale \\
\\
{\em Dipartimento di Fisica ``E.R.Caianiello'', Universit\`{a} di Salerno.}\\
{\em Istituto Nazionale di Fisica Nucleare, GC di Salerno, Italy. }}
\maketitle

\begin{abstract}
The role of the $SO(2,1)$ symmetry in General Relativity is analyzed.
Cosmological solutions of Einstein field equations invariant with respect to
a space-like Lie algebra ${\cal G}_{r}$, with $3\leq r\leq 6$ and containing 
$so(2,1)$ as a subalgebra, are also classified.

\end{abstract}

\vspace{1.5cm}

Gravitational models are usually classified in terms of their group of
isometries, namely the number of Killing vectors, the values of the
structure constants and the {\it transitivity regions, i.e.} the regions on
which the isometry group acts transitively. The classification of all
non-isomorphic isometry groups is a well known and solved issue in the
context of Group Theory; for instance, there exist $9$ non-isomorphic $3$
-dimensional groups, $G_{3}$, which yield the so-called Bianchi models.

A simpler, but important, example is provided by the $G_{2}$ groups. In this
case the corresponding Lie algebra ${\cal G}_{2}$ is described in terms of
two Killing vectors $X,Y$ satisfying the commutation relation $[X,Y]=sY$
with $s=0,1$, this corresponding respectively to the Abelian and non-Abelian
case.

Interesting gravitational fields are represented by metrics possessing a
higher degree of symmetry (for an exhaustive review on the various
classifications of such models available in the literature see for example 
\cite{KSHM79}). Among them particular attention have attracted homogeneous
or hypersurface--homogeneous models like the {\it Friedmann-Robertson-Walker
metric} ({\it FRW}) which is $G_{6}$-symmetric, the G\"{o}del metric which
is $G_{5}$-symmetric or the Kantowski-Sachs metric, $G_{4}$-symmetric.

To better understand the role of the symmetries in the classification let us
give a few definitions.

The Lie algebra of all Killing vector fields of a given metric $g$ will be
denoted by ${\cal K}${\it il}$\left( g\right) $, while {\it Killing algebra}
will denote any subalgebra ${\cal G}$ of ${\cal K}${\it il}$\left( g\right) $. 
The group corresponding to a subalgebra ${\cal G}$ of ${\cal K}${\it il}$
\left( g\right) $ is called {\it group of motions} or {\it group of
isometries }and denoted by $G_{r}$ where $r$, {\it the order of the group},
is the number of generators. If ${\cal G}={\cal K}${\it il} $\left( g\right) 
$, then the corresponding group\ is called {\it complete group of motions. }

A metric manifold is said to be {\it homogeneous} if its group of motions, $
G_{r}$, acts transitively on it, that is the whole manifold is an orbit of $
G_{r}$.

A metric manifold is said to be {\it maximally symmetric} if it has the
maximum number of Killing vector fields, {\it i.e}. if it admits a complete
group $G_{n\left( n+1\right) /2}$ of motions, where $n$ denotes the
dimension of the manifold. Of course, a maximally symmetric manifold is
homogeneous.

The {\it isotropy} or {\it stability group }of a point $p$ is a subgroup $
H_{p}$ of $G_{r}$ leaving $p$ fixed. The manifold is said to be {\it 
isotropic about} $p$ if its isotropy group is $n$-dimensional. It can be
easily shown that if the manifold is spacelike, then $H_{p}=SO\left(
n\right) $.

A manifold isotropic about every point $p$ is said to be {\it isotropic.}

It is easy to see that:

\begin{itemize}
\item  A maximally symmetric metric manifold has constant curvature. The
converse is also true.

\item  A metric manifold of constant curvature is isotropic. The converse is
also true.
\end{itemize}

Thus, a maximally symmetric manifold is isotropic and has a constant
curvature.

A {\it cosmological model} or shortly a {\it cosmology} is a Lorentzian $4$
-dimensional differential manifold foliated by $3$-dimensional submanifolds $
{\cal S}$ on which the restriction $\left. g\right| _{{\cal S}}$\ of the
metric $g$ is positive definite. These submanifolds ${\cal S}$ will be also
called {\it leaves}. A cosmology is said to be {\it homogeneous} and/or {\it 
isotropic} if the leaves ${\cal S}$ are homogeneous and/or isotropic.

The present letter moves from a previous series of articles \cite{SVV01}
where vacuum gravitational fields, invariant for a non-Abelian $2$
-dimensional Killing algebra ${\cal G}_{2}$, are exhaustively classified as
described below.

If $g$ is a metric on the space-time and ${\cal G}_{2}=span\{X,Y\}$ one of
its Killing algebras 
\begin{equation}
X,Y\in {\cal G}_{2}~~~[X,Y]=sY,~~~s=0,1\,\,\,\,,  \nonumber
\end{equation}
then the integrable involutive distribution ${\cal D}$ generated by $X$ and $
Y$ is $2$-dimensional\footnote{
The fields $X$ and $Y$, leaving invariant a metric, {\it i.e.}a symmetric
not degenerate $\left( 0,2\right) $ tensor field, cannot be parallel.}.
Moreover, if the orthogonal distribution is integrable and transversal
\footnote{
The class of metrics, so characterized, encompasses a wide variety of
gravitational models. It suffices to mention that this class includes the 
{\it Robinson-Bondi} plane-waves, the cylindrical-wave solutions, the
homogeneous cosmological models of Bianchi types I through VIII, the {\it 
pseudo-Schwarzschild} \cite{KKK99,SVV01} and {\it Kerr} solutions, the {\it 
Belinskii-Khalaktinov} \cite{BK69} general cosmological solution with a
physical singularity on portions of the so-called {\it long eras}.}, then
all the ${\cal G}_{2}$ invariant solutions of the vacuum Einstein field
equations are characterized either in terms of solutions of an algebraic
equation (the {\it tortoise equation}) or in terms of solutions of a linear
partial differential equation in the plane, depending on whether $g(Y,Y)\neq
0$ or $g(Y,Y)=0$, respectively. It was also shown that in the case $
g(Y,Y)\neq 0$ there exists a third Killing field such that ${\cal K}${\it il}
$\left( g\right) $, the complete Killing algebra, is isomorphic to $so(2,1)$
and the Killing leaves, {\it i.e.,} the transitivity regions, are $2$
-dimensional Riemann surfaces of constant curvature. Moreover, all the
invariant vacuum gravitational fields with Lorentzian signature, are locally
diffeomorphic to 
\begin{equation}
g=\frac{r-A}{r}d\tau ^{2}-\frac{r}{r-A}dr^{2}-r^{2}\left( d\vartheta
^{2}+\sinh ^{2}\vartheta d\varphi ^{2}\right)   \label{rot}
\end{equation}
where $r\in \left] 0,\infty \right[ ,\,\,\vartheta \in R,\,\,\varphi \in 
\left[ 0,2\pi \right[ $ are pseudo-spherical coordinates and $A$ an
arbitrary constant. This solution is static and $so(2,1)$ invariant; it will
be called the {\it pseudo-Schwarzschild metric} because of its resemblance
to the Schwarzschild one which is static and $so(3)$ invariant.

These results suggest to clarify the role of the $SO(2,1)$ symmetry in
General Relativity and in particular to analyze the metric 
\begin{equation}
g=dt^{2}-a^{2}(t)\left[ -\frac{dr^{2}}{kr^{2}+1}+r^{2}(d\theta ^{2}+\sinh
^{2}\theta d\phi ^{2})\right]  \label{pseudoFRW}
\end{equation}
which is similar in form to the {\it FRW} solution and invariant with
respect to the $so(2,1)$ Lie algebra 
\[
\lbrack X_{1},X_{2}]=X_{3},\quad \lbrack X_{2},X_{3}]=-X_{1},\quad \lbrack
X_{3},X_{1}]=-X_{2}, 
\]
which, in the pseudo-spherical coordinates, is spanned by 
\begin{eqnarray}
X_{1} &=&\sin \varphi \partial _{\vartheta }+\cos \varphi \coth \vartheta
\partial _{\varphi }  \nonumber \\
X_{2} &=&-\cos \varphi \partial _{\vartheta }+\sin \varphi \coth \vartheta
\partial _{\varphi }  \nonumber \\
X_{3} &=&\partial _{\varphi }.  \label{so21}
\end{eqnarray}
For $-kr^{2}>1$ the metric has Lorentzian signature and the $2$-dimensional
surfaces defined by $\left( r,t=const\right) $ may be identified with one of
the sheets of the two-sheeted space-like hyperboloid. They are also known as 
{\it pseudo-spheres}.

The pseudo-sphere \cite{BV86} is a surface with constant negative Gaussian
curvature ${\cal R}=-1/r^{2}$. It can be globally embedded in a $3$
-dimensional Minkowskian space. Let $y_{1},\,\,\,y_{2},\,\, \,y_{3}$ denote
the coordinates in the Minkowskian space, where the separation from the
origin is given by $y^{2}=-y_{1}^{2}+y_{2}^{2}+y_{3}^{2}$. These coordinates
are connected to the pseudo-spherical coordinates $\left( r,\vartheta
,\varphi \right) $ by $y_{1}=r\cosh \vartheta ,\,\,\,y_{2}=r\sinh \vartheta
\cos \varphi ,\,\,\,y_{3}=r\sinh \vartheta \sin \varphi $. The equation $
y^{2}=-r^{2}$, {\it i.e.} the locus of points equidistant from the origin,
specifies a hyperboloid of two sheets intersecting the $y_{1}$ axis at the
points $\pm r$ called {\it poles} in analogy with the sphere. Either sheet
(say the upper sheet) models an infinite spacelike surface without a
boundary; hence, the Minkowski metric becomes positive definite (Riemannian)
upon it. This surface has constant Gaussian curvature (${\cal R}=-1/r^{2}$),
and it is the only simply connected surface with this property. Other
embeddings of the pseudo-sphere in the 3-dimensional Euclidean space are
also available, for example it can be regarded as the $2$-dimensional
surface generated by the tractrix \cite{St82}, but they are not global.

The form of the metric (\ref{pseudoFRW}) is conjectured on the basis of the
well known {\it FRW} solution by essentially replacing the $SO(3)$ $2$
-dimensional orbits with those of $SO(2,1)$. Let us briefly recall the main
features of the {\it FRW} metric. In spherical coordinates ($r\in \left]
0,\infty \right[ ,\,\,\vartheta \in \left] 0,\pi \right[ ,\,\,\varphi \in 
\left[ 0,2\pi \right[ $) it has the form 
\begin{equation}
g=dt^{2}-a(t)^{2}[\frac{dr^{2}}{(1-kr^{2})}+r^{2}(d\vartheta ^{2}+\sin
^{2}\vartheta d\varphi ^{2})].  \label{FRW}
\end{equation}
It is a homogeneous and isotropic cosmology and the $3$-dimensional leaves $
{\cal S}$, defined by $t=const$, have constant scalar curvature ${\cal R}$
proportional to $k$ (${\cal R}=6k/a^{2}$). Indeed, $\dim {\cal K}${\it il}$
(g|_{{\cal S}})=6$ and the Killing vector fields $X_{i},P_{i},\{i=1,..3\}$
of $g$ 
\[
X_{i}=-\epsilon _{ijk}x_{j}\partial _{k},~~~~P_{i}=(1-\frac{kr^{2}}{4}
)\partial _{i}+\frac{1}{2}kx_{i}(r\partial _{r})
\]
close the Lie algebra 
\[
\left[ X_{i},X_{j}\right] =\sum_{k}\epsilon _{ijk}X_{k},\quad \left[
P_{i},P_{j}\right] =k\sum_{k}\epsilon _{ijk}X_{k},\quad \left[ X_{i},P_{j}
\right] =\sum_{k}\epsilon _{ijk}Y_{j}.
\]
Here $\epsilon _{ijk}$ is the Levi-Civita tensor density and the standard
relation between spherical and Cartesian coordinates, $x_{i}$, is
understood. For positive $k$ the six vectors span the Lie algebra of $SO(4)$
, for $k=0$ they span the Lie algebra of the semidirect product $SO(3){
\times }^{\prime }R^{3}$, whereas $k$ negative corresponds to the proper
Lorentz group $SO(3,1)$. Since the maximally symmetric and isotropic leaves $
{\cal S}$ are spacelike, the isotropy group is for all the three cases $SO(3)
$. In order to evidentiate the special role of $SO(3)$ and because of some
parallelism we may trace for $so(2,1)$ invariant situations, it is useful to
rewrite the three Lie algebras respectively as $so(3)\oplus so(3)$, $\
so(3)\oplus ^{\prime }R^{3}$, $so(3)\bar{\oplus}sb(2,C)$, where $\oplus
^{\prime }$ denotes the semidirect sum and $\bar{\oplus}$ denotes a fully
non-commutative sum of two Lie algebras which act nontrivially on each
other by
adjoint action (more details on this structure are given later on); $sb(2,C)$
is the Lie algebra of type V in the Bianchi classification.

Going back to the metric (\ref{pseudoFRW}), the question arises whether it
admits other Killing fields and, if this is the case, what is the complete
Killing algebra. Solving the Killing equations 
\begin{equation}
P^{\alpha }\partial _{\alpha }g_{\mu \nu }+\partial _{\mu }P^{\alpha
}g_{\alpha \nu }+\partial _{\nu }P^{\alpha }g_{\mu \alpha }=0
\label{killingeq}
\end{equation}
we find three more Killing fields, 
\begin{eqnarray}
P_{1} &=&b(r)\left[ \sinh \vartheta \cos \varphi \partial _{r}-\frac{\cos
\varphi \cosh \vartheta }{r}\partial _{\vartheta }+\frac{\sin \varphi }{
r\sinh \vartheta }\partial _{\varphi }\right]  \nonumber \\
P_{2} &=&b(r)\left[ \sinh \vartheta \sin \varphi \partial _{r}-\frac{\sin
\varphi \cosh \vartheta }{r}\partial _{\vartheta }-\frac{\cos \varphi }{
r\sinh \vartheta }\partial _{\varphi }\right]  \nonumber \\
P_{3} &=&b(r)\left[ \cosh \vartheta \partial _{r}-\frac{\sinh \vartheta }{r}
\partial _{\vartheta }\right]  \label{so21pi}
\end{eqnarray}
with $b(r)=\sqrt{1-kr^{2}}$. Together with the vector fields $X_{i}$ of Eqs.
(\ref{so21}) they span the Lie algebra 
\begin{equation}
\lbrack
X_{i},X_{j}]=c_{ijk}X_{k},~~~[X_{i},P_{j}]=c_{ijk}P_{k},~~~[P_{i},P_{j}]=kc_{ijk}X_{k}.
\label{kilalg}
\end{equation}
According to the value of $k$ three different situations occur. It can be
easily checked that for $k=0$, ${\cal K}il\left( g\right) $ is the
semidirect sum $so(2,1)\oplus ^{\prime }R^{3}$ while for positive $k$, $
{\cal K}il\left( g\right) $ is the direct sum $so(2,1)\oplus so(2,1)$, that
is to say $so(2,2)$. Both the choices yield a metric of indefinite
signature. The case $k$ negative and $-kr^{2}>1$ is the interesting one from
a cosmological point of view. Here the signature of the metric is Lorentzian
and, performing the change of basis 
\begin{eqnarray*}
Y_{1} &=&{\frac{P_{1}}{\sqrt{-k}}},\,\,Y_{2}={\frac{P_{2}}{\sqrt{-k}}},\,\
Y_{3}=-X_{3}, \\
Q_{1} &=&X_{1},\,\,Q_{2}=X_{2},\,\,Q_{3}={\frac{P_{3}}{\sqrt{-k}}}
\end{eqnarray*}
to reach a more conventional form, we discover that the Killing algebra (\ref
{kilalg}) is nothing but the Lorentz algebra $so(3,1)$. The scalar curvature
of the spacelike submanifolds ${\cal S}$ is constant and negative 
\begin{equation}
{\cal R}_{({\cal S})}=6\frac{k}{a^{2}}~.  \label{curv}
\end{equation}
Therefore, we conclude that the metric (\ref{pseudoFRW}) with $k$ negative
is diffeomorphic to the {\it FRW} metric of negative curvature.

To complete the analysis let us investigate the other cosmological solutions
of the Einstein field equations, invariant with respect to a spacelike
algebra ${\cal G}_{r}$, with $3\leq r\leq 6$, containing $so(2,1)$ as a
subalgebra.

The most general $so(2,1)$ invariant metric can be locally written in
pseudospherical coordinates as 
\begin{equation}
g=A(r,t)dt^{2}-B(r,t)dr^{2}-F(r,t)r^{2}(d\vartheta ^{2}+H(\vartheta
)d\varphi ^{2})  \label{metric}
\end{equation}
where $H(\vartheta )$ is one of the functions $\sinh ^{2}\vartheta $ or $
-\cosh ^{2}\vartheta $. With positive definite functions $A,B,F$ and $
H(\vartheta )=\sinh ^{2}\vartheta $ the metric is Lorentzian and the $2$
-dimensional surfaces defined by $\left( r,t=const\right) $ are
pseudo-spheres, that is, space-like surfaces of constant negative curvature.
In the other case, corresponding to $-\cosh ^{2}\vartheta $, the $2$
-dimensional surfaces defined by $\left( r,t=const\right) $ are one-sheeted
time-like hyperboloids and will not be considered here. As already mentioned
the pseudo-Schwarzschild metric (\ref{rot}) is locally the only $so\left(
2,1\right) $-invariant solution of the Einstein field equations in the
vacuum \footnote{
This yields an extension of the Birkoff theorem to the $so(2,1)$ invariant
case.}. This is the static solution found in \cite{SVV01} and also, in the
context of warped solutions, in \cite{KKK99}.

The more physically interesting gravitational field, generated by a
distribution of matter described by an energy-momentum tensor $T_{\mu \nu }$
and reducing in the vacuum to the previous one, is given by 
\begin{equation}
g=f(r)dt^{2}-h(r)dr^{2}-r^{2}(d\vartheta ^{2}+\sinh ^{2}\vartheta d\varphi
^{2}),  \label{smetric}
\end{equation}
where the $so(2,1)$-invariant positive functions $f(r)$ and $h(r)$ satisfy
the equations 
\begin{eqnarray*}
8\pi T_{00} &=&h^{\prime }(rh^{2})^{-1}+\frac{1}{r^{2}}(1-h^{-1}) \\
8\pi T_{11} &=&f^{\prime }(rhf)^{-1}-\frac{1}{r^{2}}(1-h^{-1}) \\
8\pi T_{22} &=&{\frac{1}{2}}f^{\prime }(rhf)^{-1}-{\frac{1}{2}}h^{\prime
}(rh^{2})^{-1}+{\frac{1}{2}}(fh)^{-1/2}[(fh)^{-1/2}f^{\prime }]^{\prime },
\end{eqnarray*}
the apex denoting the derivation with respect to $r$.

For a perfect fluid, with energy momentum tensor field of the type $T_{\mu
\nu }=\rho u_{\mu }u_{\nu }+P(g_{\mu \nu }+u_{\mu }u_{\nu })$, the above
equations give 
\begin{eqnarray*}
h &=&\left( 1-2m(r)/r\right) ^{-1}, \\
\frac{d\psi }{dr} &=&\frac{4\pi Pr^{3}+m}{r(r-2m)} \\
\frac{dP}{dr} &=&-(P+\rho )\frac{4\pi Pr^{3}+m}{r(r-2m)}
\end{eqnarray*}
where $\psi =\ln \sqrt{f}$ and $m(r)=4\pi \int_{1}^{r}\rho (r)r^{2}dr+C$,
the constant $C=h(1)^{-1}-1$ being determined by the boundary conditions.

The quantity $m(r)$, when integrated over the whole volume of the source,
represents the total mass. Unlike the $so\left( 3\right) $ invariant compact
case, the hyperbolic symmetry gives to the source an infinite extension.
Then, a constant density is not allowed whereas each function of $r$
decreasing faster than $1/r^{3}$ will do.

\begin{itemize}
\item  {\bf ${\cal G}_{3}=so\left( 2,1\right) $ invariant cosmologies}

The most general $so\left( 2,1\right) $ invariant cosmology is given by the
metric (\ref{metric}) with $A(r,t)=1$: 
\begin{equation}
g=dt^{2}-B(r,t)dr^{2}-F(r,t)r^{2}(d\vartheta ^{2}+\sinh ^{2}\vartheta
d\varphi ^{2})  \label{nmetric}
\end{equation}
Cosmologies with maximal symmetry $SO(2,1)$ are not homogeneous since the
orbits of $SO(2,1)$ are $2$-dimensional. They belong to the class of Bianchi
cosmologies \cite{EM69}. An interesting point of view which deserves more
investigation is to regard them as limiting cases of models with higher
symmetries in the presence of sources which retain only the $SO(2,1)$
symmetry.

\item  {\bf ${\cal G}_{4}$ invariant cosmologies}

An additional space-like Killing field for the above metric has to commute
with the $so(2,1)$ generators thus yielding a central extension of $so(2,1)$
and will have the general form 
\begin{equation}
P=P(r,t)\frac{\partial }{\partial r}.  \nonumber
\end{equation}
The Killing equations (\ref{killingeq}) are satisfied by 
\begin{equation}
P=P(r)\frac{\partial }{\partial r}{,\,\,\,\,\,\,}g=-dt^{2}+\frac{B(t)}{
P(r)^{2}}dr^{2}+F(t)(d\vartheta ^{2}+\sinh ^{2}\vartheta d\varphi ^{2}), 
\nonumber
\end{equation}
where $P(r){\,\,}$\ and${\,\,\,\,}B(t)$ are arbitrary non vanishing positive
functions and the time coordinate $t$ has been rescaled by $\sqrt{A\left(
t\right) }$. By defining a new radial coordinate, this gravitational field
may be reexpressed as 
\begin{equation}
g=-dt^{2}+b(t)dr^{2}+f(t)(d\vartheta ^{2}+\sinh ^{2}\vartheta d\varphi ^{2}).
\label{pseudokantowski}
\end{equation}
In this form it is immediately recognized as one of the Kantowski--Sachs
cosmological solutions\footnote{
Among them the most known is the one invariant with respect to {\bf ${\cal G}
_{4}$}$=so(3)\oplus R$ obtained from the previous one by replacing the $
SO(2,1)$ orbits with those of $SO(3)$.} \cite{KS66}. The metric (\ref
{pseudokantowski}) is spatially homogeneous as the leaves ${\cal S}$ are
orbits of $G_{4}$, but not isotropic, $SO(2,1)$ being only a global symmetry 
\cite{KS66,KSHM79}.

\item  {\bf ${\cal G}_{5}$ invariant cosmologies}

There are no ${\cal G}_{5}$ invariant cosmologies because a $3$-dimensional
metric manifold cannot admit a complete group $G_{5}$ of motions \cite{Fu03}.

The G\"{o}del model \cite{St82} 
\begin{equation}
g=dx^{2}+e^{2x}dy^{2}/2+dz^{2}-(e^{x}dy+cdt)^{2},  \nonumber
\end{equation}
with $\dim {\cal K}${\it il}$\left( g\right) =5$ and Killing fields 
\begin{eqnarray*}
X_{1} &=&\frac{\partial }{\partial y},\,\ X_{2}=\frac{\partial }{\partial x}
-y\frac{\partial }{\partial y},\,\,X_{3}=y\frac{\partial }{\partial x}
+(e^{-2x}-\frac{1}{2}y^{2})\frac{\partial }{\partial y}-2e^{-x}\frac{
\partial }{\partial t}, \\
Y_{1} &=&\frac{\partial }{\partial z},\,\,Y_{2}=\frac{\partial }{\partial t}
\end{eqnarray*}
is a static solution whose isometry group acts transitively on the whole
space-time which is then a homogeneous manifold. The Killing fields $Y_{i}$
commute between them and the vector fields $X_{i}$ close the $so\left(
2,1\right) $ Lie algebra, but do not have space-like orbits.

\item  {\bf ${\cal G}_{6}$ invariant cosmologies}

In this case, the leaves ${\cal S}$ are maximally symmetric and then
isotropic and of constant curvature. Hence we search for three additional
Killing fields $P_{i}$ which, together with the $so(2,1)$ generators, span
the Lie algebra of any possible $G_{6}$ with space-like $3$-dimensional
orbits. Apart from $SO(2,1)$, because of the isotropy, it must contain $SO(3)
$ as a subgroup. On the other hand, maximally symmetric manifolds are
uniquely specified by the numbers of eigenvalues of the metric that are
positive or negative and by the sign of the scalar curvature \cite{Ei26,We72}
; it follows that the $G_{6}$ group we are looking for is the proper Lorentz
group $SO(3,1)$ and the corresponding invariant cosmology has to be
diffeomorphic to the {\it FRW} cosmology (\ref{FRW}) with negative spatial
curvature, no matter what cumbersome coordinate system is adopted. Then, the
only solution is represented by the metric in Eq. (\ref{pseudoFRW}). Let us
see how it can be derived in a more systematic manner.

The most general ${\cal G}_{6}$ including $so(2,1)$ as a subalgebra may be
given the Lie algebra structure \cite{Gi74} 
\begin{eqnarray}
\lbrack X_{i},X_{j}] &=&c_{ijk}X_{k}  \label{nso21} \\
\lbrack X_{i},P_{j}] &=&\epsilon _{1}c_{ijk}P_{k}+\epsilon _{2}f_{ijk}X_{k}
\label{der} \\
\lbrack P_{i},P_{j}] &=&\epsilon _{3}c_{ijk}X_{k}+\epsilon _{4}f_{ijk}P_{k}
\label{plie}
\end{eqnarray}
where $c_{ijk}$ are the structure constants of $so(2,1)$, $\ f_{ijk}$ are
some unknown structure constants and $\epsilon _{i}=0,\pm 1$, all of them
constrained by the Jacobi identity. Up to isomorphisms we have the following
cases.

\begin{itemize}
\item[i)]  $\epsilon _{1}=1,~~\epsilon _{2}=\epsilon _{4}=0,~~\epsilon
_{3}=0,\pm 1$. This corresponds to search $G_{6}$ in the form of a principal
fibre bundle having $SO(2,1)$ as structure group;

\item[ii)]  $\epsilon _{1}=\epsilon _{2}=\epsilon _{3}=0,~~\epsilon _{4}=1$.
This is the direct sum, $so(2,1)\oplus {\cal G}_{3}$, which can be
considered for any of the $9$ Bianchi types. Imposing the Killing equations
to the generators of ${\cal G}_{3}$ we find no solutions of this form, if $
so(2,1)$ is realized as in (\ref{so21});

\item[iii)]  $\epsilon _{1}=\epsilon _{4}=1,~~\epsilon _{3}=0$. This is the
sum of two $3$-dimensional Lie algebras not commuting between them, one of
them being $so(2,1)$. The case $\epsilon _{2}=0$ corresponds to a semidirect
sum, whereas $\epsilon _{2}=1$ yields a fully non-commutative sum of Lie
algebras. In this case $so(2,1)$ is said to be a {\it Lie bialgebra} \cite
{Dr83}. The compatibility condition for two Lie algebras to be given such a
structure is 
\begin{equation}
c_{ijk}f_{krs}+c_{irk}f_{jks}-f_{irk}c_{kjs}-c_{jrk}f_{iks}+f_{jrk}c_{kis}=0~.
\label{comp}
\end{equation}
There are only two solutions for this equation, one with $f_{ijk}$ all
vanishing which yields the semidirect sum $so(2,1)\oplus ^{\prime }R^{3}$
(this is the only semidirect sum compatible with the $so(2,1)$ structure),
the other one with $f_{ijk}$ given by the structure constants of $sb(2,C)$,
the Lie algebra of type $V$ in the Bianchi classification. They both
correspond to an indefinite metric; the first one is rediscovered below in
the context of case (i).
\end{itemize}

Let us analyze the case (i) in more detail. From the condition (\ref{der}),
with $X_{i}\in so(2,1)$ the fields $P_{i}$ must be of the form (\ref{so21pi}
) with $b(r)$ an arbitrary function of $r$. The condition (\ref{plie}) fixes
the $r$ dependence to be $b(r)=(-\epsilon _{3}r^{2}+C)^{1/2}$, with $C$ an
arbitrary constant. Moreover, the Killing equations for the metric (\ref
{nmetric}) restricted to the leaves ${\cal S}$, give $B(r)=\frac{B}{
(-\epsilon _{3}r^{2}+C)},\,\,\,\,F(r)=-\frac{B}{C},$ with $B$ an arbitrary
constant. Thus, the restricted metric takes the form 
\begin{equation}
\left. g\right| _{{\cal S}}=-\frac{dr^{2}}{1+\epsilon _{3}r^{2}/B}
+r^{2}(d\vartheta ^{2}+\sinh ^{2}\vartheta d\varphi ^{2}),  \label{g3metric}
\end{equation}
where the coordinate $r$ has been rescaled by the factor of $\sqrt{-B/C}$.
The space-time metric is then 
\begin{equation}
g=dt^{2}-a^{2}(t)[-\frac{dr^{2}}{1+\epsilon _{3}r^{2}/B}+r^{2}(d\vartheta
^{2}+\sinh ^{2}\vartheta d\varphi ^{2})].  \label{FRWneg}
\end{equation}
This can be seen to coincide with the solution (\ref{pseudoFRW}) with $
\epsilon _{3}/B=k$. Hence, according to the value of $\epsilon _{3}/B$ the
three situations already described occur.

To summarize, we have found yet an alternative form of the {\it FRW}
metric, for the case of negative scalar curvature, in terms of the symmetry
subgroup $SO(2,1)$. This corresponds to the fact that both $SO(3)$ and $
SO(2,1)$ are subgroups of the Lorentz group, so that we can choose to adapt
our system of coordinates to either one or the other. With our choice the
leaves ${\cal S}$ are foliated by pseudospheres, {\it i.e} space-like orbits
of $SO(2,1)$ which are non-compact and of constant negative curvature. The
conclusion is that the $SO(2,1)$ invariance is much more stringent than the $
SO(3)$ invariance, because there is only one possible ${\cal G}_{6}$
-invariant cosmology admitting $SO(2,1)$ as a subgroup and this is the well
known Lorentz invariant, negative curvature, {\it FRW} metric.
\end{itemize}

\section*{Acknowledgments}

The authors wish to thank G. Bimonte, G. Marmo and G.Sparano for interesting
remarks and the Erwin Schr\"{o}dinger Institut in Vienna for the kind
hospitality. The work was supported in part by the {\it Progetto di Ricerca
di Interesse Nazionale} SINTESI 2000.

\end{document}